# New Results for the Heterogeneous Multi-Processor Scheduling Problem using a Fast, Effective Local Search and Random Disruption

John Levine[1], Graeme Ritchie[2], Alastair Andrew[1] and Simon Gates[3]

[1] Department of Computer and Information Sciences,
  University of Strathclyde, Glasgow, UK

[2] European Bioinformatics Institute,
  Wellcome Trust Genome Campus,
  Hinxton, Cambridge, UK

[3] Clinical Trials Unit,
  Warwick Medical School,
  University of Warwick, Coventry, UK

**Abstract**

The efficient scheduling of independent computational tasks in a heterogeneous computing environment is an important problem that occurs in domains such as Grid and Cloud computing. Finding optimal schedules is an NP-hard problem in general, so we have to rely on approximate algorithms to come up schedules that are as near to optimal as possible. In our previous work on this problem, we applied a fast, effective local search to generate reasonably good schedules in a short amount of time and used ant colony optimisation (ACO) to incrementally improve those schedules over a longer time period. In this work, we replace the ACO component with a random disruption algorithm and find that this produces results which are competitive with the current state of the art over a 90 second execution time. We also ran our algorithm for a longer time period on 12 well-known benchmark instances and as a result provide new upper bounds for these instances.

**Introduction**

The heterogeneous multi-processor scheduling problem (MPS) is concerned with allocating a set of independent tasks onto a set of heterogeneous processors that are capable of completing the tasks. An optimal solution to the problem is one that allocates each task to a processor such that the total time to complete all the tasks (i.e. the makespan) is minimised.

This scheduling problem is important in modern computing environments due to the need to spread computational load across a wide variety of computing devices. This problem occurs most notably in Grid and Cloud computing where multiple end users want to submit their computational tasks to an abstract device called "the Grid" or "the Cloud" which is in reality made up of lots of different, interconnected computing devices.

The version of the problem we are tackling was introduced by Braun *et al.* (2001). Their formulation of the problem is as follows: you have a set of tasks, $T$, and a set of processors, $P$. Typically, $T$ will be an order of magnitude larger than $P$ (more tasks than processors). The problem is expressed as an expected-time-to-compute (ETC) matrix: for each $t \in T$ and for each $p \in P$, the value of $ETC(t, p)$ is the amount of time that we expect processor $p$ will take to complete task $t$. A solution to the problem consists of finding a complete allocation of all tasks such that each task is executed on a single processor and the total time to execute all the tasks, as calculated from the ETC matrix, is minimised. This problem is known to be NP-hard as it is a generalised reformulation of problem SS8 from Garey and Johnson (1979).

We used two approaches in our previous work on this problem. In the first (Ritchie and Levine, 2003) we used a fast, effective local search on the schedules produced by a greedy construction algorithm and found that this produced very competitive results compared with the results given in Braun *et al.* (2001). In the second approach (Ritchie and Levine, 2004) we used Ant Colony Optimisation combined with local search to produce an anytime algorithm that could incrementally improve an initial schedule produced by a greedy construction heuristic.

Based on our observations with these methods, we now present an algorithm for tackling the problem that is both simpler than the ACO method and produces results that are currently the best known for these problem instances. We use the fast local search from our earlier paper combined with a technique known as "shaking" – applying random disruption to a solution that has reached a local optimum. We compare our results with the current best reported by Nesmachnow *et al.* (2012), who use a multi-population micro genetic algorithm combined with local search. Because the instances used vary greatly in the magnitude of their optimal makespan, we use the Wilcoxon matched-pairs signed ranks test to compare the results. This test shows that the results of the local search and shaking algorithm are significantly better than those reported by Nesmachnow *et al.*

Nesmachnow *et al.* also give lower bounds on the solutions for the Braun *et al.* problems, by relaxing the problem and using linear programming to form a solution. We investigated how close our local search algorithm could get to these lower bounds by running our algorithm for several weeks (over the Christmas period, when the computers were less busy). The local search algorithm found new optima for all 12 Braun et al. problems, with the new solutions being only a fraction above the lower bounds reported by Nesmachnow.

**Algorithm**

The algorithm we use in this work can be described as follows:

The first phase of the algorithm uses a greedy construction heuristic to create an initial solution to work from. In the experiments described below we use the min-min heuristic presented by Braun *et al.* which we also used in our earlier work. This heuristic works by establishing the minimum completion time for every unscheduled job (given the current schedule) and then assigns the job that extends the makespan of the current schedule the least. The intuition is that this will keep the load on the processors balanced throughout the allocation.

The second phase of the algorithm consists of running the local search algorithm described by Ritchie and Levine (2003). This works by first finding the "problem processor" (i.e. the one whose tasks will run for the longest time and which is therefore responsible for the value of the current makespan). The local search operates by considering a neighbourhood consisting of (i) swapping a single task on this processor with a task allocated to some other processor in the schedule, plus (ii) transferring a single task on this processor onto another processor in the schedule. The neighbourhood is searched exhaustively and the solution with the best improvement is selected. The whole process is then repeated, with a new problem processor being identified, until no further improvement is possible.

The third phase of the algorithm is the random disruption or "shaking" phase. This is engaged when a solution has reached a local optimum, i.e. all of the solutions in the neighbourhood of the solution are non-improving. The idea of disruption is to try to change the solution just enough to move it out of the attraction of the local optimum just found, but while staying close enough to the current optimum to ensure that the disrupted solution retains some elements of the best solution found so far. The intuition here is that keeping a fairly high percentage of the solution intact should result in a solution that can climb the "neighbouring hillsides" around the local optimum just found. The disruption used here consists of choosing a random number of swaps to apply to the solution (up to a given limit – here set to be 9 as this was found in initial trials to give the best results).

Finally, after the disruption phase has (hopefully) knocked the solution away from the local optimum, the second phase of the algorithm (local search) is run again to identify a new local optimum. If the makespan of this new solution is equal to or lower than the makespan of the previous optimum, then it becomes the new optimum and the disruption phase is then applied. If the new local optimum is worse than the old one, then the new optimum is discarded and the random disruption phase is applied to the old optimum again.

**Instances**

In order to simulate various possible heterogeneous scheduling problems as realistically as possible, Braun *et al.* (2001) define different types of ETC matrix according to three metrics: task heterogeneity, machine heterogeneity and consistency.

The task heterogeneity is defined as the amount of variance possible among the execution times of the jobs. Two possible values were defined: high and low. Machine heterogeneity, on the other hand, represents the possible variation of the running time of a particular job across all the processors, and again has two values: high and low.

In order to try to capture some other possible features of real scheduling problems, three different ETC consistencies were used: consistent, inconsistent and semi-consistent. An ETC matrix is said to be consistent if whenever a processor $p_j$ executes a task $t_i$ faster than another processor $p_k$ then $p_j$ will execute *all* other jobs faster than $p_k$. A consistent ETC matrix can therefore be seen as modelling a heterogeneous system in which the processors differ only in their processing speed. In an inconsistent ETC a processor $p_j$ may execute some tasks faster then $p_k$ and some slower. An inconsistent ETC matrix could therefore simulate a network in which there are different types of machine available, e.g. a UNIX machine may perform jobs that involve a lot of symbolic computation faster than a Windows machine, but will perform jobs that involve a lot of floating point arithmetic slower. A semi-consistent ETC matrix is an inconsistent matrix which has a consistent sub-matrix of a predefined size, and so could simulate, for example, a computational grid which incorporates a sub-network of similar UNIX machines (but with different processor speeds), but also includes an array of different computational devices.

These different considerations combine to leave us with 12 distinct types of possible ETC matrix (e.g. high task, low machine heterogeneity in an inconsistent matrix, etc.) which simulate a range of different possible heterogeneous systems. The matrices used in the comparison study of (Braun et al., 2001) were randomly generated with various constraints to attempt to simulate each of the matrix types described above as realistically as possible. Braun *et al.* provide 100

ETC matrices for each of the 12 classes; in keeping with other work, we use only the first ETC matrix from each class to facilitate exact comparison of results. These 12 instances have names of the form w_x_yyzz.0 where:

w denotes the probability distribution used to generate the ETC matrices:
– only uniform distributions were used so this is u for all files.
x denotes the type of consistency, one of:
– c: consistent matrix
– i: inconsistent matrix
– s: semi-consistent
yy denotes the task heterogeneity, one of:
– hi: high heterogeneity
– lo: low heterogeneity
zz denotes machine heterogeneity, one of
– hi: high heterogeneity
– lo: low heterogeneity

Thus (for example) the instance `u_c_lohi.0` was generated using a uniform distribution, is a consistent problem (as defined above) and has a low value of task heterogeneity but a high value for the machine heterogeneity.

**Experiments**

The main aim of the experiments was to compare our results with the current best results for these instances which are reported by Nesmachnow *et al.* (2012). The algorithm they use is a multi-population genetic algorithm which runs on a parallel cluster containing 16 CPUs for 90 seconds of run time. Our algorithm is single-threaded, so in order to simulate the result we would have obtained on parallel hardware, we ran each instance 16 times for 90 seconds and then took the best of these 16 results. Each algorithm was run 50 times on each instance and the mean and standard deviation of the makespan are reported.

In a second experiment, we ran our algorithm on the 12 instances and left it running over the Christmas period (two weeks) when the machines were not in use for other activities. We were interested in a number of questions: would the local search with disruption get stuck, or would it continue to find improvements? How close to the theoretical lower bounds for the instances, as calculated by Nesmachnow *et al.* using linear programming, would it be possible to get? And would there be any variation in solvability across the 12 different instances?

**Results**

The results for the first experiment are shown in Table 1. The results from Nesmachnow et al.'s parallel micro evolutionary algorithm are shown in columns 2 and 3, with the results for our local search and disruption algorithm shown in columns 4 and 5. To compare the 12 results, we applied the Wilcoxon matched-pairs signed ranks test, which uses the size of the differences in the makespans between the two algorithms, so is better than a simple sign test (which would just use whichever was the "winner" in each problem). Using this test, we obtained a p-value close to 0, showing that the difference between the two sets of results is very unlikely to have been obtained by chance. Nesmachnow's results were the best reported on these instances until this paper, so this shows that the relatively simple technique of local search and shaking can produce excellent results on this problem.

|            | Parallel micro EA |         | Local search + shaking |         |
|------------|-------------------|---------|------------------------|---------|
|            | Mean              | Std dev | Mean                   | Std dev |
| `u_c_hihi.0` | **7394702.7**   | 0.09%   | 7401477.4              | 0.06%   |
| `u_c_hilo.0` | 153193.7        | 0.04%   | **153156.2**           | 0.02%   |
| `u_c_lohi.0` | **239706.2**    | 0.08%   | 239859.3               | 0.06%   |
| `u_c_lolo.0` | 5152.3          | 0.04%   | **5146.5**             | 0.02%   |
| `u_i_hihi.0` | 2947896.4       | 0.14%   | **2932328.8**          | 0.07%   |
| `u_i_hilo.0` | 73531.4         | 0.10%   | **73310.9**            | 0.03%   |
| `u_i_lohi.0` | 102402.8        | 0.17%   | **101685.4**           | 0.06%   |
| `u_i_lolo.0` | 2547.1          | 0.09%   | **2539.6**             | 0.04%   |
| `u_s_hihi.0` | 4123537.3       | 0.27%   | **4095948.4**          | 0.09%   |
| `u_s_hilo.0` | 96020.5         | 0.10%   | **95820.2**            | 0.03%   |
| `u_s_lohi.0` | 122744.4        | 0.23%   | **121734.2**           | 0.13%   |
| `u_s_lolo.0` | 3438.3          | 0.07%   | **3427.3**             | 0.03%   |

Table 1: Comparison of the results obtained by Nesmachnow et al.'s parallel micro evolutionary algorithm of with the results from local search plus random disruption on the 12 instances from Braun et al. (2001).

The results from the second experiment, where we let our algorithm run for approximately two weeks on each instance are shown in Table 2, alongside the theoretical lower bounds given by Nesmachnow *et al.* using linear programming. It was notable that these solutions continued to improve in quality over the entire runtime (i.e. the algorithm didn't get stuck) although towards the end of the runs the improvements were infrequent and very small.

| Problem      | LS + shaking | Lower bound | Gap to LB |
|--------------|--------------|-------------|-----------|
| `u_c_hihi.0` | 7360142.1    | 7346524.2   | 0.19%     |
| `u_c_hilo.0` | 152815.4     | 152700.4    | 0.08%     |
| `u_c_lohi.0` | 238768.4     | 238138.1    | 0.26%     |
| `u_c_lolo.0` | 5137.9       | 5132.8      | 0.10%     |
| `u_i_hihi.0` | 2930069.0    | 2909326.6   | 0.71%     |
| `u_i_hilo.0` | 73182.6      | 73057.9     | 0.17%     |
| `u_i_lohi.0` | 101547.1     | 101063.4    | 0.48%     |
| `u_i_lolo.0` | 2536.1       | 2529.0      | 0.28%     |
| `u_s_hihi.0` | 4087295.7    | 4063563.7   | 0.58%     |
| `u_s_hilo.0` | 95584.0      | 95419.0     | 0.17%     |
| `u_s_lohi.0` | 121147.6     | 120452.3    | 0.57%     |
| `u_s_lolo.0` | 3420.8       | 3414.8      | 0.18%     |

Table 2: extended runs of the local search and shaking algorithm on the 12 instances from Braun *et al.* (2001) together with the theoretical lower bounds from Nesmachnow *et al.* (2012)

**Conclusions and Further Work**

We were genuinely surprised that the simple algorithm reported here could outperform both ant colony optimisation and a parallel micro evolutionary algorithm on this problem. It would appear that a single point search is both sufficient to produce good quality schedules for this problem and that the search landscapes for this problem are well suited for this algorithm, with the search managing to continue to improve upon solutions even after many days of CPU time.

The next phase of this research will involve trying to use a finer grained evaluation function for solutions rather than just the makespan: two solutions may have the same makespan (complete in the same amount of time) but the one that has more "spare capacity" in it would be preferred as it offers more opportunities for local search to do good things (i.e. shift tasks around in order to lower the overall makespan).